\documentclass[aps,showpacs,twocolumn]{revtex4}
\usepackage{amsmath}
\input{epsf.tex}
\epsfverbosetrue

\begin{document}

\input epsf
\newcommand{\infig}[2]{\centerline{\epsfxsize=#2\columnwidth \epsfbox{#1}}}
\newcommand{\be}{\begin{equation}}
\newcommand{\nn}{\nonumber}
\newcommand{\ee}{\end{equation}}
\newcommand{\bea}{\begin{eqnarray}}
\newcommand{\eea}{\end{eqnarray}}
\newcommand{\bin}[2]{\left(\begin{array}{c} #1 \\ #2\end{array}\right)}
\newcommand{\pred}{^{\mbox{\small{p}}}}
\newcommand{\retr}{^{\mbox{\small{r}}}}
\newcommand{\retd}{_{\mbox{\small{r}}}}
\newcommand{\p}{_{\mbox{\small{p}}}}
\newcommand{\m}{_{\mbox{\small{m}}}}
\newcommand{\tr}{\mbox{Tr}}
\newcommand{\rs}[1]{_{\mbox{\small{#1}}}}	
\newcommand{\ru}[1]{^{\mbox{\small{#1}}}}
\newcommand{\ris}[2]{_{\mbox{\small{#1}}{#2}}}	
\title{Preamplified photodetectors for high-fidelity postselecting optical 
devices}

\author{John Jeffers}

\affiliation{SUPA, Department of Physics, University of Strathclyde, John 
Anderson Building, 107 Rottenrow, Glasgow G4 0NG, U.K.}

\begin{abstract}
The fidelity of postselecting devices based on direct photon number detection 
can be significantly improved by insertion of a phase-insensitive optical 
amplifier in front of the detector. The scheme is simple, and the cost to the 
probability of obtaining the appropriate detector outcome is low.
\end{abstract} 
\maketitle

\section{Introduction}
\begin{figure}[h]
\infig{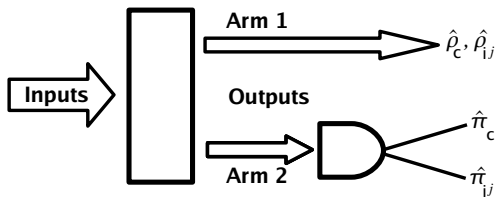} {0.75} \caption{\protect\small A general postselecting device.  
\label{postselector}}
\end{figure}

The reliable production of tailored quantum states is one of the main 
challenges in quantum information. Experiments in optics rely on postselection 
to do this \cite{Bachor,pdc,scissors}. A general perfect postselecting 
device is shown in Fig. \ref{postselector}. A multi-component quantum state is 
fed into a device which transforms the input state. One of the outputs (arm 2) 
is measured, and when the measurement gives a particular result (represented 
mathematically by a probability operator measure element $\hat{\pi}\rs{c}$) the 
device produces the correct useful state $\hat{\rho}\rs{c}$ in arm 1. If one of 
a set of incorrect measurement results ($\hat{\pi}\ris{i}{j}$) is found the 
device produces an incorrect state $\hat{\rho}\ris{i}{j}$.

Often in optics the unselected state is mixed, so postselection is essential. 
Simple examples include the generation of heralded single photon states from 
twin-beam parametric downconversion \cite{pdc}, or the quantum scissors device 
\cite{scissors}, which produces a superposition of zero and one photon. 
Interest in the topic received a further boost with the realisation that it 
might be possible to perform scalable linear optical quantum computing, with 
both state production and logic gate operation relying on postselection 
\cite{KLM,Ralph,Kok}. Variations on this include the one-way quantum computer, 
which produces states from measurements performed on cluster states 
\cite{Nielsen}.

Imperfections, either in internal device components or in detection, are a 
serious problem for real postselectors. As a result they do not make the state 
that they would have made if they had been functioning perfectly.  
Photodetectors are poor at present, suffering from low quantum efficiency, from 
dark counts and from an inability to distinguish between higher photon numbers. 
This means that the single photon number readout from a detector, which ought 
to correspond to a pure number-state measurement, in fact corresponds to a 
mixed measured state \cite{mixed}. The effect of this on the postselector 
output state is to mix $\hat{\rho}\rs{c}$ with the set of incorrect states 
$\hat{\rho}\ris{i}{j}$. 

The quality of device output can be quantified by the fidelity, which is a 
measure of the closeness of the imperfect output state produced, 
$\hat{\rho}\rs{c}^\prime$, to the required output state, $\hat{\rho}\rs{c}$. 
The standard definition, if $\hat{\rho}\rs{c}$ is pure, is 
$F=\tr \left( \hat{\rho}\rs{c} \hat{\rho}\rs{c}^\prime \right)$, which is unity 
for a perfect device \cite{Jozsa}. 

Optical quantum information processors, which might be composed of thousands of 
postselectors, require extremely high fidelities. The improvement of 
photodetectors to such levels is unrealistic, but there have been proposals for 
novel detection schemes to improve fidelity. For example, photon-added homodyne 
detection \cite{Branc} does this, but with a high cost to the probability of 
detecting the required state (reduced by more than 100-fold if the fidelity is 
to be greater than 0.99). Another proposed approach is to use a nonlinear 
optical material formed by atoms in a dielectric waveguide to perform quantum 
nondemolition detection of photon number \cite{Munro}. This scheme has the 
appealing feature that the detected photons can be reused. Despite this, both 
schemes are much more complicated than direct detection, and the former 
requires photon number states as a resource.

Here a scheme is proposed which uses inefficient direct photodetection preceded 
by a phase-insensitive optical amplifier. Amplifiers are not typically used in 
quantum optical experiments as they add noise photons \cite{Caves}, which swamp 
any quantum characteristics of the amplified state \cite{Mandel}. They have 
been used to offset detector inefficiency, improving the signal to noise ratio 
for direct detection autocorrelation measurements of laser light \cite{harris}, 
but never for quantum states. However, there have been recent improvements to 
amplifiers for quantum systems \cite{amps}. Also, it has been shown, using 
retrodictive quantum theory \cite{retro}, that the state transformation made by 
an amplifier of gain $G$ forwards in time is the same as that made by an 
attenuator with transmission $1/G$ backwards in time and {\it vice versa} 
\cite{ampatt}, which leads to seemingly strange input photon number expectation 
values \cite{JLJ}.

In the next section the retrodictive conditional probability is introduced as a 
measure of fidelity appropriate for optical postselecting devices. Then results 
are provided for postselection based on recording zero or one photocount, 
followed by a simple explanantion. An analysis of the cost of amplification 
follows, in terms of a reduction in the photocount probability. In the final 
section the results are discussed and conclusions presented. 

\section{Device fidelity based on preamplified detection}
\subsection{Conditional probability as a fidelity measure}
For a pure state postselector with perfect internal components, based on 
detecting photon number states, substitution of $\hat{\rho}\rs{c}$ and 
$\hat{\rho}\rs{c}^\prime$ allows the fidelity to be expressed as a sum of terms 
\cite{JJ}
\bea
F = P\retr(\mbox{c}|\mbox{c}) + \sum_j P\retr(\mbox{i}j|\mbox{c}) 
\tr_1[\hat{\rho}\rs{c}\hat{\rho}\ris{i}{j}].
\eea
Here $P\retr(\mbox{c}|\mbox{c})$ is the retrodictive conditional probability 
that the number of photons in the measurement arm (2) is the same as that 
indicated by the detector, and $P\retr(\mbox{i}j|\mbox{c})$ is one of the set 
of probabilities that the number of photons is different from that indicated by 
the detector. The first term, which we denote $F\retd$, has been proposed as a 
simple measure of fidelity \cite{JJ}. It has advantages over $F$. Firstly it 
depends only on measurement arm properties: the `prior' probabilities 
\cite{peggjeff} of detectable states in the measurement arm and the properties 
of the mixing performed there. Secondly, it is the natural quantity to maximise 
in order to enhance fidelity (if $F\retd$ is unity the confidence in the 
detector result is perfect, as is $F$) \cite{Croke}. Often the overlaps between 
the correct output state and the incorrect ones will be small. Furthermore, as 
the detection scheme is improved, the probabilities that the measured state is 
not that indicated by the detector diminish. Thus $F\retd$ forms a close lower 
bound on $F$. 

\begin{figure}[h]
\infig{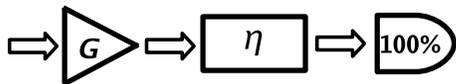} {0.7} \caption{\protect\small An amplifier precedes the 
imperfect detector, modeled by an attenuator in front of a perfect detector.  
\label{ampdet}}
\end{figure}

The detector is characterised by an efficiency $\eta$, and discriminates 
between different photon numbers. This is not typical, but detectors which 
discriminate between zero, 1 and more than 1 photon exist \cite{Rohide}, and 
here postselection based solely on recording zero or 1 count is considered, as 
this is the most prevalent. For practical purposes, then the detector is 
equivalent to a perfect discriminating device preceded by an attenuator of 
transmission $\eta$ \cite{Shapiro}. The detection system is completed by an 
ideal amplifier of gain $G$ (Fig. \ref{ampdet}) which adds the minimum amount 
of noise. Discussion of both extra amplifier noise and dark counts is postponed 
until later. The fictitious attenuator and real amplifier jointly form a 
compound mixing device preceding the perfect detector \cite{JJ}. 

Retrodictive conditional probabilities can be found from Bayes' Theorem. If the 
postselecting device is supposed to produce the correct state when $n$ photons 
are detected then $F\retd$ is
\bea
F\retd(n) = \frac {p_n P\pred(n|n)}{\sum_m p_m P\pred(n|m)},
\eea
where $p_m$ is the prior probability of $m$ photons in the measurement arm, and 
$P\pred(n|m)$ is the predictive conditional probability that $n$ photons exit 
the compound mixing device and are recorded as counts at the perfect detector, 
given that $m$ enter it. The required probabilities can be straightforwardly 
calculated. The prior probabilities are the diagonal elements, in the photon 
number basis, of the arm 2 state formed by tracing the joint output state of 
the two arms over arm 1. The conditional probabilities are well-known from the 
quantum theories of the amplifier and attenuator \cite{JLJ,Shepherd},
\bea
P\pred(n|m)=\sum_{q=n}^\infty \binom{q}{n} \eta^n (1-\eta)^{q-n} 
\binom{q}{m} \frac{(G-1)^{q-m}}{G^{q+1}}.
\eea
The denominator in $F\retd(n)$ is the probability that $n$ counts are recorded, 
which is also altered by the amplifier. 

\subsection{Postselection based on zero or one photocount}
First postselection based on recording 0 photocounts is examined. The prior 
photon number probability distribution will be fixed by the particular device 
under consideration, but a distribution must be chosen for calculation purposes.
\begin{figure}[h]
\infig{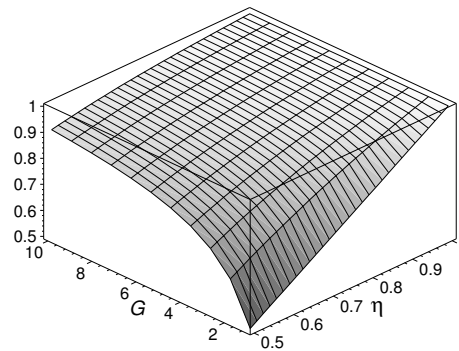} {0.7} \caption{\protect\small Fidelity against $G$ and $\eta$ 
for postselection based on 0 counts, for each photon number in the measurement 
arm having equal prior probability.  \label{fig3}}
\end{figure}
If all of the prior photon number probabilities are equal Fig. \ref{fig3} is 
obtained, which shows $F\retd(0)$ as a function of $\eta$ and $G$. The greater 
the gain, the greater is the improvement over the no-amplifier, $G=1$ limit 
\cite{JJLconf}. The device will show increase in fidelity as the gain increases 
for any ($p_0 \neq 0$) prior probability distribution. A simple example might 
be a two-photon state generator formed by a single photon input into each input 
arm of a  50/50 beam splitter. If no photons are detected in one output arm 
then two are produced in the other. For this device all of the prior 
probabilities vanish except $p_0=p_2=1/2$. 
\begin{figure}[h]
\infig{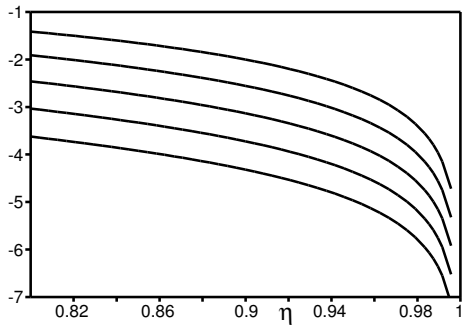} {0.7} \caption{\protect\small $\log _{10}(1-F\retd(0))$ against 
$\eta$ for the two-photon state generator. The lines are for (from top to 
bottom) $G=$1, 2, 4, 8 and 16. \label{fig4}}
\end{figure}
A plot of the log of the difference between the fidelity and 1 against $\eta$ 
is shown in Fig. \ref{fig4} for various amplifier gains. Here as $G$ is 
increased the fidelity tends to 1 even more rapidly, a function of the 
particular prior probability distribution. It is possible to reach extremely 
high fidelities even for relatively modest gain.

\begin{figure}[h]
\infig{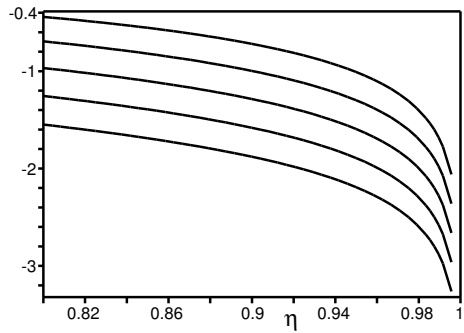} {0.7} \caption{\protect\small $\log _{10}(1-F\retd(1))$ against 
$\eta$ for equal prior probability of each nonzero photon number in the 
measurement arm. The lines are for (from top to bottom) $G=$1, 2, 4, 8 and 16. 
\label{fig5}}
\end{figure}

For postselection based on one count, with equal prior photon number 
probabilities, fidelity can be increased for low $G$ only if $\eta$ is below 
about 0.7. The situation is different, however, if the measurement arm contains 
at least one photon, which is illustrated by Fig. \ref{fig5}. 
Increasing the gain increases the fidelity to arbitrarily close to unity. Any 
prior distribution for which $p_0$ vanishes will show improvement in fidelity 
as the gain increases for all values of $\eta$.

\subsection{Simple explanations for fidelity increase}
A clarification of the physics behind the effect is found by considering the 
measurement arm states corresponding to the measurement results. For a perfect 
detector these are the pure states indicated by the detector, but for an 
imperfect detector they are mixed \cite{mixed}. 
\begin{figure}[h]
\infig{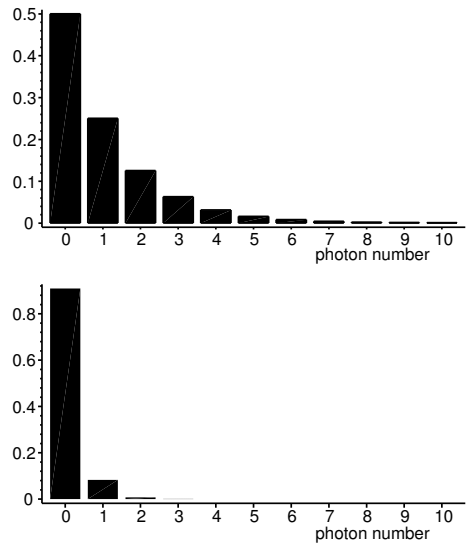} {0.7} \caption{\protect\small Photon number probability 
histograms for detection of 0 photons by a detector with $\eta=0.5$. The upper 
histogram is the measured state if there is no amplifier present, and the lower 
one is for $G=10$. \label{fig6}}
\end{figure}
Fig. \ref{fig6} shows histograms of the photon number probability distribution 
of the mixed state corresponding to detection of zero photons by an imperfect 
detector. The effect of the amplifier (lower histogram) is to `attenuate' the 
state such that the mean photon number in the projected state is $1/G$ times 
the mean photon number in the projected state with no amplifier present (upper 
histogram) \cite{ampatt,JLJ}. This causes probability to `pile up' at lower 
photon numbers, and in particular on zero. 

The same effect is responsible for fidelity increase based on 1 count. For high 
enough gain the most likely photon number, if 1 is counted, is zero, then 1, 
then 2 etc. The prior distribution then becomes important. Excluding the 
possibility of zero photons entering the detector amounts to omitting the zero 
photon component of the projected state and then renormalizing, so that 1 
photon is the lowest possible photon number in the distribution. Otherwise the 
fidelity decreases with increasing $G$ for high $\eta$.

An alternative view is that the amplifier shifts the photon number of the 
$n$-photon component of its input (the state represented by the prior 
distribution) from exactly $n$ to a mean of $(n+1)G-1$. In other words it 
separates photon numbers by a factor $G$ and adds $G-1$ photons (although the 
random nature of the process means that there is some overlap between the 
shifted distributions for different initial photon numbers). For reasonable 
values of $\eta$ it is very unlikely that zero counts will be obtained from a 
shifted 1 or 2 photon state, as it is very unlikely that so many photons will 
be lost at the attenuator. If zero counts are obtained, it is therefore more 
likely that no photon was present.

\subsection{Photocount probability cost}
\begin{figure}[h]
\infig{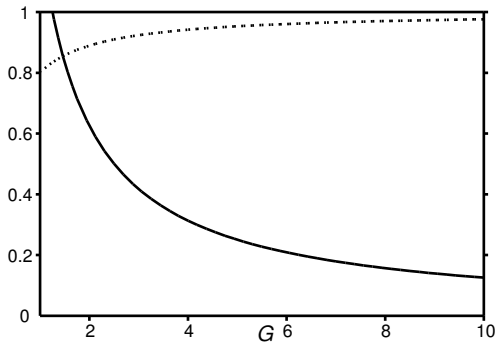} {0.75} \caption{\protect\small Relative probability 
$P(0)/P(0,\eta=G=1)$ of no counts being recorded at the detector as a function 
of $G$ for $\eta=0.8$. Also plotted (dashed) is the fidelity.\label{fig7}}
\end{figure}

There is a cost associated with the large fidelity increases which are possible 
using preamplification, and this is seen in the photocount probabilities. The 
ratio of the probability that the detector records zero counts to this same 
probability for perfect detection quantifies this cost, and is shown in 
Fig. \ref{fig7} (for equal prior probabilities).
The curve is insensitive to $\eta$. The relative probability is reduced, but 
not excessively so. Even for $G=10$ the probability is only reduced to 
1/8$^{th}$ of its value for perfect detection. The fidelity obtained for 
$\eta=0.8, G=10$ is 0.975, but this is for a flat prior photon number 
distrubution (the corresponding figure for $\eta=0.9$ is close to 0.99). For 
states with distributions with small or vanishing probabilities of higher 
photon numbers the improvement can be vast, as Fig. \ref{fig4} shows.

\section{Conclusions}
In this paper it has been shown that an optical preamplifier can significantly 
improve the fidelity of postselectors based solely on imperfect direct 
photodetection. A further advantage is that the amplifier does not drastically 
decrease the probability of device operation. Even this modest decrease in 
probability could be offset by including other detection results as a signature 
of the required number of photons in the measurement arm ({\it e.g.} if 
$G\eta \gg 1$ and one count is obtained the most likely photon number in the 
measurement arm is zero).

The scheme works best if the detected photon number is the minimum number 
possible in the measurement arm. Thus it is especially useful for improving the 
fidelity based on detecting zero photons. For detecting single photons the 
improvement is almost as great, but the benefit of the method rests on the 
ability to produce measurement arm states which do not contain a significant 
vacuum component. This is a matter of postselector design and photon production 
technique, which is rapidly improving under the impetus of the quantum 
information challenge \cite{Matsu,Geremia}.  For postselectors with detections 
in more than one output arm, such as the quantum scissors \cite{scissors} 
amplifiers can be placed in front of each detector, and similar improvements in 
fidelity can be found. 

Up to this point the amplifier has been assumed to add the minimum number of 
noise photons. The main effect of extra noise photons is to decrease the 
fidelity obtained for a particular gain. As was stated earlier the amplifier 
shifts and separates different input photon numbers, and spreads the output 
distributions. Extra noise spreads the distributions more, so that different 
photon number components of the input are not so distinguishable. This decrease 
in fidelity can sometimes be partially offset by increasing $G$, or the effect 
itself may be small because of the particular prior photon number distribution 
in the measurement arm. One might think that detector dark counts would cause a 
similar decrease in fidelity but this is not the case. Fidelity based on zero 
counts is unaffected by dark counts (no counts obtained $\implies$ no dark 
counts obtained). Fidelity based on 1 count can be improved slightly by a 
nonzero dark count rate. These effects will be explored more fully in later 
work.

Optical amplifiers are overlooked in quantum information experiments, largely 
because of the necessary added noise photons, but it should be noted that these 
`noise' photons are indistinguishable from amplified signal photons. Provided 
that the noise photons added into non signal modes can be excluded from the 
detection process, the photons added in the signal mode can play a useful role, 
and can sometimes be regarded as an additive component to the multiplicative 
gain $G$. For direct detection in postselectors both the quantum nature and the 
phase of the detected signal state are unimportant, but the result of the 
detection process and the confidence in that result are paramount. Amplifiers 
will not help the first two quantities, but can significantly improve the last.

\section*{Acknowledgments}
The author thanks the Engineering and Physical Sciences Research Council for 
financial support and Steve Barnett and Rodney Loudon for useful discussions.

\end{document}